\documentclass{article}

\usepackage{PRIMEarxiv}

\usepackage[utf8]{inputenc} % allow utf-8 input
\usepackage[T1]{fontenc}    % use 8-bit T1 fonts
\usepackage{hyperref}       % hyperlinks
\usepackage{url}            % simple URL typesetting
\usepackage{booktabs}       % professional-quality tables
\usepackage{amsfonts}       % blackboard math symbols
\usepackage{nicefrac}       % compact symbols for 1/2, etc.
\usepackage{microtype}      % microtypography
\usepackage{lipsum}
\usepackage{fancyhdr}       % header
\usepackage{graphicx}       % graphics
\graphicspath{{media/}}     % organize your images and other figures under media/ folder

\title{20736-node Weighted Max-Cut Problem Solving by Quadrature Photonic Spatial Ising Machine
\thanks{} 
}

\author{
  Xin Ye, Wenjia Zhang, Shaomeng Wang, Xiaoxuan Yang, Zuyuan He \\
  State Key Laboratory of Advanced Optical Communication Systems and Networks \\
  Shanghai Jiao Tong University \\
  Shanghai 200240, China\\
  \texttt{\{Wenjia Zhang\}wenjia.zhang@sjtu.edu.cn} \\
}

\begin{document}
\maketitle

\begin{abstract}
To tackle challenging combinatorial optimization problems, analog computing machines based on the nature-inspired Ising model are attracting increasing attentions in order to disruptively overcome the impending limitations on conventional electronic computers.
Photonic spatial Ising machine has become an unique and primitive solution with all-to-all connections to solve large-scale Max-cut problems. However, spin configuration and flipping requires two independent sets of spatial light modulators (SLMs) for amplitude and phase modulation, which will lead to tremendous engineering difficulty of optical alignment and coupling.
We report a novel quadrature photonic spatial-Euler Ising machine to realize large-scale and flexible spin-interaction configuration and spin-flip in a single spatial light modulator, and develop a noise enhancement approach by adding
digital white noise onto detected optical signals.
We experimentally show that such proposal accelerates solving (un)weighted, (non)fully connected, 20736-node Max-cut problems, which offers obvious advantages over simulation and heuristic algorithm results in digital computers.
\end{abstract}

\section{Introduction}
The Max-Cut problem is one of fundamental NP-hard problems in combinatorial optimization, which can describe the practical applications such as date clustering, machine scheduling and image recognition. This problem can be formulated into an equivalent Ising model without local fields, given by $H=-\sum_{<l,k>}J_{l,k}x_{l}x_{k}$ where $J_{l,k}$ is the interaction between spins and binary spin state $x_{l} \in \{1,-1 \}$. And photonic Ising machines are designed to search for ground state of the Ising model by either iterative sampling or directly evolving the ensemble energy regarding the established mapping of a particular combinatorial problem \cite{hon,pi,ruan,Sun}. Among these solutions, spatial photonic Ising machines encoding the spins as a phase matrix in spatial light modulators (SLMs), translate the Ising model into another form as 
\begin{equation}
H=-\sum_{<l,k>}\varepsilon_{l}\varepsilon_{k}x_{l}x_{k}
    \label{H1}
\end{equation}
in which the interaction coefficient $J_{l,k}$ is set by the amplitude modulation $\varepsilon_{l}$ and $\varepsilon_{k}$ \cite{pi}. This scheme is highly suitable for solving large-scale Max-Cut problems that exhibit fully connected interactions, owing to its superior connectivity and scalability \cite{ruan}. However, several proposed Ising machines tend to focus on solving the benchmark unweighted Max-Cut problems, as they can be easily mapped onto the Ising model with $J_{l,k}$ taking values of $\{0, \pm1\}$ while the weighted ones are more practical and complex.

In this paper, we successfully solves 20736-node weighted Max-Cut problems with the quadrature photonic spatial Ising machine(Q-SIM). To configure the weights, we perform intensity configuration based on Euler's Formula by extending the quadrature phase configuration in previously proposed architecture \cite{Sun}. 
Furthermore, we extend our experiments to instances of dense graphs and compare our results with those obtained through numerical simulations and other reviewed methods. Our experimental results show a 33\% improvement in the maximum cut value over the simulation results and a 34\% improvement over the standard Sahni-Gonzales (SG) algorithm. Moreover, our approach provides a substantial overall speed-up.

\section{Architecture and principle}
\begin{figure*}[t]
\centering  %插入的图片居中表示
\includegraphics[width=0.8\linewidth] {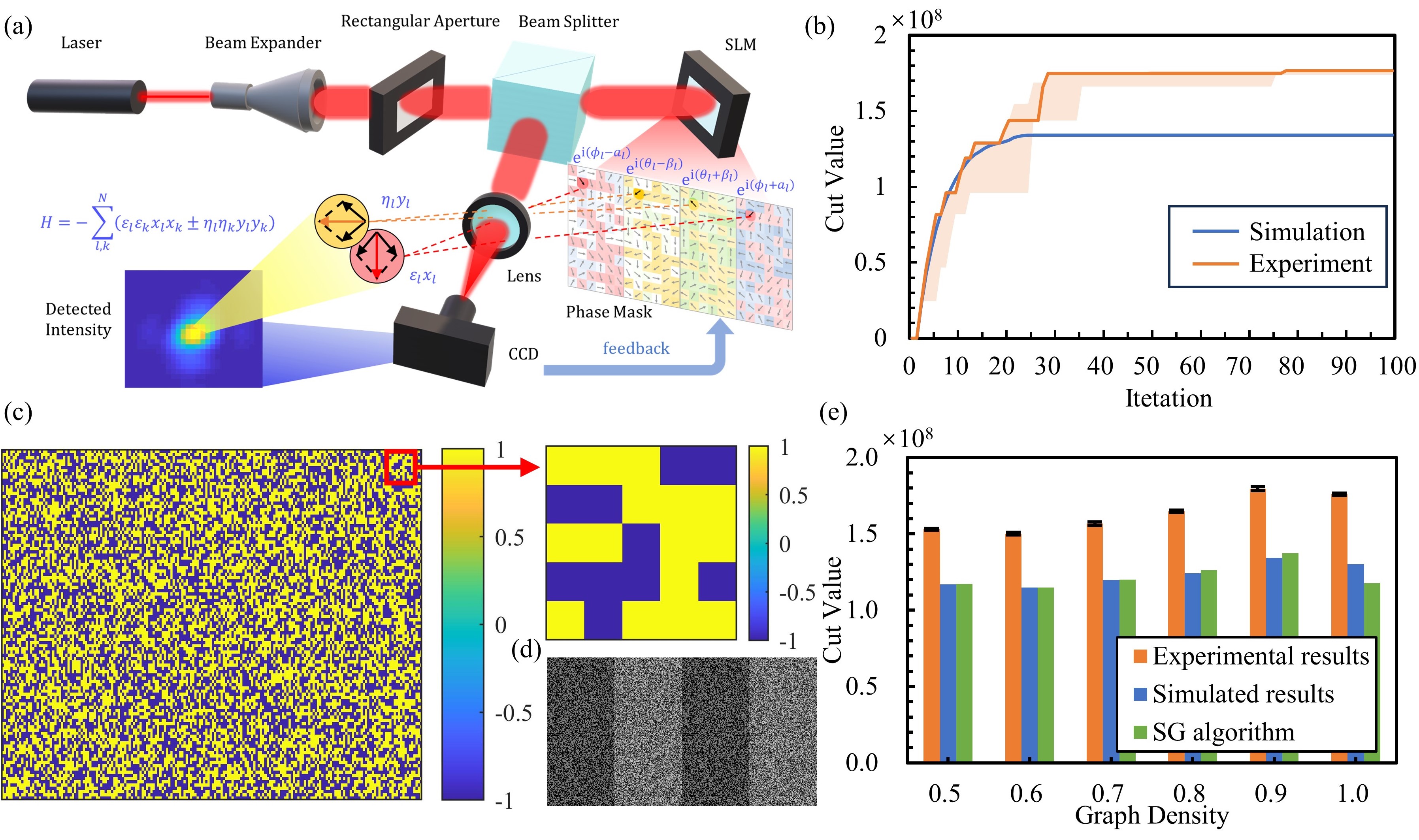}   
\caption{
(a) The schematic and principle of Q-SIM with phase-to-amplitude configuration. 
(b) Experimental (orange line) and simulated (blue line) evolution of the cut value. The light orange region represents the interval of the result distribution from the ten experiments, with the mean indicated by the orange line.
(c) Final experimental spin states in (b).
(d) Final experimental phase mask in (b).
(e) Experimental results for Max-Cut problems with graph densities of $[0.5,1.0]$.
}
\label{CouplingPrinciple}
\end{figure*}

As shown in Fig.\ref{CouplingPrinciple}(a), an extended coherent light source shines on the SLM screen. The phase mask of SLM is composed of four parts that encode both the interaction coefficients and the spin states, allowing for a spin with amplitude information to be represented by four distinct components: $e^{i(\phi_{k}-\alpha_{k})}$, $e^{i(\theta_{k}-\beta_{k})}$, $e^{i(\phi_{k}+\alpha_{k})}$, $e^{i(\theta_{k}+\beta_{k})}$. Here, $\phi_{k}\in\{0,\pi\}$ and $\theta_{k}\in\{\frac{\pi}{2},-\frac{\pi}{2}\}$ are two sets of mutually orthogonal phases to construct the Q-SIM, as demonstrated in our previous work\cite{Sun}. These orthogonal phases allow for flexible configurations of low-rank matrices, overcoming the inherent restriction of rank=1 imposed by the form of Eq.\ref{H1}. In addition, some non-fully-connected Ising models with negative amplitudes are successfully configured with this architecture. On the other hand, the sum of $e^{i(\phi_{k}-\alpha_{k})}$ and $e^{i(\phi_{k}+\alpha_{k})}$ can form the amplitude of the spin, according to Euler's formula
\begin{equation}
    \varepsilon_{k}x_{k}=\frac{1}{2}[e^{i(\phi_{k}-\alpha_{k})}+e^{i(\phi_{k}+\alpha_{k})}]
    \label{Phase and amplitude}
\end{equation}
where $\varepsilon_{k}=\cos\alpha_{k}$. This approach simplifies the step of amplitude configuration while eliminating the limitation of non-negative amplitude. Then, after the two-dimensional Fourier transform by the lens, the central intensity detected by the CCD Camera is given in the form of
\begin{equation}
    I(0,0)=(x^{T}\varepsilon+y^{T}\eta)(\overline{\varepsilon^{T} x+\eta^{T}y })
    \label{intensity}
\end{equation}
A specific derivation procedure regarding the principle of the quadrature photonic spatial Ising machine and the intensity configuration method based on Euler's formula is provided in Appendices A and B.

Now the weighted Max-Cut instances can be mapped to the Ising model for solution. When a cut is defined, the corresponding cut value can be written as $W=\frac{1}{2}\sum_{<l,k>}w_{l,k}(1-x_{l}x_{k})$, in which $w_{l,k}$ is the edge weight between  the l-th node and the k-th node. The related Hamiltonian we use is $H=\sum_{<l,k>}w_{l,k}x_{l}x_{k}$ and the weight can be expressed as $w_{l,k}=\cos\alpha_{l}\cos\alpha_{k}\pm\cos\beta_{l}\cos\beta_{k}$. Hence we can search for the maximum cut value by maximizing the central intensity during the experiment. Here, we calculate the Euclidean distance $\|I_{T}-I\|_{2}$ between the initial detected intensity$I$ and target intensity$I_{T}$ as a cost function of the simulated annealing (SA) algorithm, thus generating a new phase mask to refresh SLM screen. This procedure is continuously cycled to govern the Hamiltonian evolution until the system stabilizes to the ground state. Details about the experimental setup and the operational procedures based on the SA algorithm are demonstrated in Appendices C and D.

\section{Experiments and Discussion}

Given that the exact solvers generally fail with 1000 nodes, we need to perform a reference calculation on the conventional electrical computing platform. To tackle larger instances, we opt to utilize a classical greedy heuristic algorithm known as the SG algorithm \cite{hon}.
On average, it requires 11 hours to execute the SG algorithm on CPUs (Intel i9-13900K, 5.8 GHz) for addressing the Max-Cut problems of 20736 nodes. In the case of an all-to-all Max-Cut problem, the SG method generates a maximum cut value of $1.178\times10^8$, while our method produces a higher cut value of $1.759\times10^8$ with a 122 times speedup, as shown in Fig. \ref{CouplingPrinciple} (b). Additionally, we conduct simulations to emulate the functioning of the photonic Ising machine. Notably, our findings reveals that the simulation results are also inferior to the experimental results.

Finally, we extrapolate our experiments and simulations for the Max-Cut problem with graph densities of $[0.5,1.0]$ compared with the SG algorithm. The results are shown as Fig. \ref{CouplingPrinciple} (e), which statistically demonstrate that our scheme offers compelling advantages for handling large-scale Max-Cut problem that outweighs electronic computers, in comparison with both simulations and the SG algorithm. The experimental maximum cut values exceed the SG algorithm by an average of 34\% and achieve a maximum of 49\% with the graph density of 1.0, which precisely captures the inherent advantage of fully connected systems. Additionally, the experimental results routinely outperform the simulated results by roughly 33\%. These will be illustrated in the computational results given in Appendix E.
We speculate that the detection susceptible to noise may cause some perturbation in experiments, making it easier to jump out of the local energy minima. This intrinsic property fits better with the SA algorithm and thus improves the machine performance. In fact, a related work reported that noise-enhanced photonic Ising machines can be used to solve large-scale combinatorial optimization problems\cite{pi2}.

\section{Discussion and conclusion}
We conduct a comparative analysis of our system performance, as elaborated in Appendix F, in solving Max-Cut problems against other Ising machines and identify the following advantages: (1) Efficient resolution of large-scale Max-Cut problems. (2) Flexible mapping capabilities for (non)fully connected Max-Cut problems, allowing for arbitrary amplitude assignments. (3) An uncomplicated and cost-effective experimental setup. 

In summary, we performed extensive experiments of the Q-SIM with the phase-to-amplitude configuration applied to the instance of the Max-Cut problem. This system effectively addresses weighted problems with dense graphs of up to 20736 nodes, resulting in over 30\% enhancements compared to classical solvers. 
Consequently, our proposal showcases superior optimization performance and rapid computational speed within the optical computing paradigm, making it a highly competitive solution for addressing large-scale NP problems.

\section*{Acknowledgments}
This work is supported by the National Key Research and Development Program of China under Grant 2019YFB1802903, National Natural Science Foundation of China under Grant 62175146 and 62235011 and Major Key Project of PCL (PCL2021A14).

%Bibliography
\bibliographystyle{unsrt}  
\bibliography{EuerSIMArxiv}  

\begin{appendix}

\section{Demonstration of the quadrature photonic spatial Ising machine}
In a spatial photonic Ising machine, the electric field after amplitude modulation and phase encoding is
\begin{equation}
E_{k}=\sum_{k}^{N}\varepsilon_{k}x_{k}rect_{W}(x-x_{k})
\label{ek}
\end{equation}
where $x_{k}=e^{i\phi_{k}}\in\{-1,1\}$. The electric field  at the back focal plane of the lens corresponds to the Fourier transform of $E_{k}$. 
\begin{equation}
\widetilde{E_{k}}=\sum_{k}^{N}\varepsilon_{k}x_{k}\delta_{W}(x-x_{k})
\label{ek_fft}
\end{equation}
Therefore, the central intensity detected by the CCD is
\begin{equation}
    I(0,0)=x^{T}\varepsilon^{T}\varepsilon x
    \label{intensity1}
\end{equation}
where $x^{N}\in\{-1,1\}^N$.

In the quadrature photonic spatial Ising Machine, the SLM is configured with two simultaneous Ising models of the same scale. The spin values ($x^{N}, y^{N}$) are updated simultaneously and determined by a relationship matrix A
\begin{equation}
    y=Ax
    \label{y=Ax}
\end{equation}
where $A=diag(a_{1},a_{2},...,a_{N})$, $a_{k}\in\{-i,i\}$. The two sets of phases encoded on the SLM are mutually orthogonal: $\phi_{k}(x)\in\{0,\pi\}$, $\phi_{k}(y)\in\{\frac{\pi}{2},-\frac{\pi}{2}\}$.
Similar to Eq.\ref{intensity1}, the detected intensity image on CCD is
\begin{equation}
    I(0,0)=(x^{T}\varepsilon^{T}+y^{T}\eta^{T})(\overline{\varepsilon x+\eta y })
    \label{intensity2}
\end{equation}
Thus, the interaction matrix is given in the form of
\begin{equation}
    J=\varepsilon^{T}\varepsilon\pm\eta^{T} \eta
    \label{J}
\end{equation}
In this equation, the combination of free parameters $\varepsilon_{k}$ and $\eta_{k}$ determines the value of $J_{k}$ as positive or negative, achieving the flexible configuration of the low-rank matrix.

\section{Detailed deduction of amplitude configuration method based on Euler's formula}
A beam of incident light after amplitude modulation shines onto the SLM, and its transverse electric field can be expressed as Eq.\ref{ek}.
%\begin{equation}
%E_{k}=\sum_{k}^{N}\varepsilon_{k}x_{k}rect_{W}(x-x_{k})
%\label{1}
%\end{equation}
%where $x_{k}=e^{i\phi_{k}}\in\{-1,1\}$.
As a corollary to Euler's formula $e^{i\cos^{-1}\varepsilon'}+e^{-i\cos^{-1}\varepsilon'}=2\varepsilon'$, let $\alpha=\cos^{-1}\varepsilon'$, then the amplitude normalized to $[-1,1]$ is expressed as
\begin{equation}
\varepsilon'=\frac{e^{i\alpha}+e^{-i\alpha}}{2}
\label{2}
\end{equation}
Substituting Eq.\ref{2} in Eq.\ref{ek}, we have 
\begin{equation}
E_{k}=\sum_{k}^{N}\frac{1}{2}[e^{i(\phi_{k}-\alpha_{k})}+e^{i(\phi_{k}+\alpha_{k})}]rect_{W}(x-x_{k})
\label{3}
\end{equation}
Eq.\ref{3} can be rewritten as
 \begin{equation}
E_{k}=\frac{1}{2}\sum_{k}^{N}[e^{i(\phi_{k}+(-1)^{2k-1}\alpha_{k})}+e^{i(\phi_{k}+(-1)^{2k}\alpha_{k})}]rect_{W}(x-x_{k})
\label{4}
\end{equation}
The set $\{l|l=2k-1,k=1,2...N\}$ and $\{l|l=2k,k=1,2...N\}$ can be combined into a single set $\{l|l=1,2...2N\}$, so Eq.\ref{4} can be simplified as 
 \begin{equation}
E_{l}=\frac{1}{2}\sum_{l}^{2N}[e^{i(\phi_{l}+(-1)^{l}\alpha_{l})}]rect_{W}(x-x_{l})
\label{5}
\end{equation}
Eq.\ref{5} satisfies the form of the electric field after phase modulation, eliminating the requirement of amplitude modulation. To do so, twice the number of pixels is required and the phase on the SLM $\phi_{l,SLM}=\phi_{l}+(-1)^{l}\alpha_{l}$.

\section{Experimental setup}
 To be specific, we use a stabilized red HeNe laser (Thorlabs' Stabilized Helium Neon Laser) with a center wavelength of 632 nm as the light source. When the laser operates in intensity-stabilized mode, its output power remains constant at 1.2 mW with an intensity stabilization of ±0.2\%, thus avoiding the degradation of power fluctuation on the experimental results. A 40X beam expansion and a rectangular aperture are arranged sequentially behind the laser to generate a rectangular spot large enough to cover the employed reflective phase-only SLM (HOLOEYE LETO-3-CFS-127, 1920×1080 pixels, pixel pitch of 6.4 µm) entirely. Therefore, all the pixels are activated to maximize pixel utilization while minimizing modulation-related pixel alignment issues. Additionally, we chose 10×10 pixels as a spin unit for uniform modulation in order to reduce position deviation. To separate the incident and reflected optical paths of SLM, we place a beam splitter (BS) in the optical path. A lens (focal length $f=150$ mm) placed at one arm of the BS performs a 2D Fourier transform of the optical field from the SLM. In the rear, the charge-coupled device (CCD, QSI 660i, 2758×2208 pixels, 16bit digital resolution) camera detects the intensity of the light at the back focal plane of the lens. Theoretically, the light intensity at the center location comprises sufficient information for the measurement. However, the single-pixel light intensity is susceptible to fatal errors caused by positional deviation and a full-pixel image would contain more noise coming from higher-order diffracted light as well as natural light. Hence, we finalized a detection area of 10 × 10 pixels with its 4.54 µm pixel. Moreover, we detect 5 times in each iteration and average the results, with a one-second time interval between the two detections.

\section{Algorithm and operation}
\begin{figure}[t]
\includegraphics[width=1\linewidth]{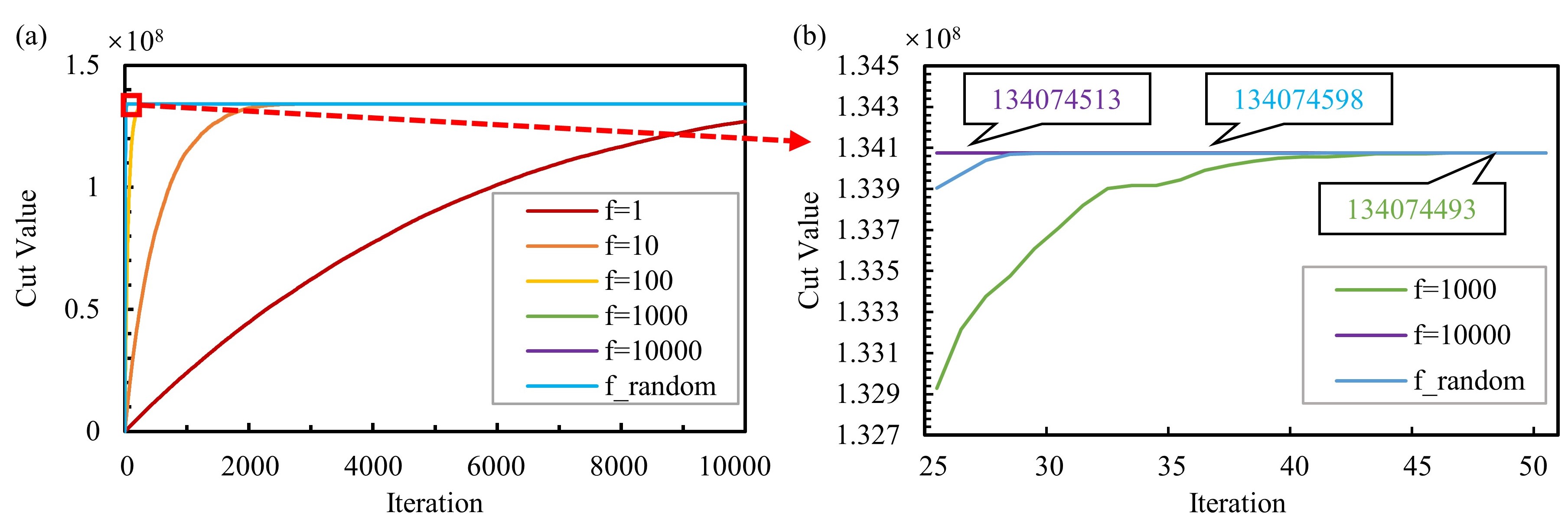}
\centering
\caption{
\textbf{ The cut value of 20736-node Max-Cut problem as a function of the number of spins flipped per iteration.} 
(a) Simulated evolution of the cut value at some specific numbers of spins flipped per iteration $(f={1, 10, 100, 1000, 10000})$ and random numbers $(f\_random)$.
(b) Details on the optimizing results between the large fixed numbers and random numbers of flips.
}
\label{fig:1}
\end{figure}
Using the SA algorithm, we search the solution for Max-Cut problem through the following steps.

Input: spin number $N$, weighted vector $w$, n×n Ising interaction matrix $J=w^{T}*w$, annealing coefficient $\lambda$, end temperature $T_{end}$, target intensity $I_{T}$

Output: optimal spin vector $x_{end}\in\{-1,1\}^N$, maximum cut value $W_{end}$

Initialization: random spin vector $x \gets x_{0}\in\{-1,1\}^N$, cut value $W \gets W_{0}$, Hamiltonian $H \gets H_{0}$, annealing temperature $T \gets T_{0}$
 
While $T > T_{end}$
 
 \quad Randomly flip several spins to generate the new spin state;
 
 \quad Update SLM;
 
 \quad Detect the intensity $I_{new}$ by CCD;

 \quad Calculate $W_{new}$, $H_{new}$;
 
 \quad If $\|I_{T}-I_{new}\|_{2}<\|I_{T}-I\|_{2}$
 
 \qquad $x\gets x_{new}$;
  
 \qquad $W\gets W_{new}$;
  
 \qquad $H\gets H_{new}$;

 \quad  Else
  
  \qquad If $rand(0,1) < e^{-\frac{H-H_{new}}{T}}$
  
  \quad \qquad  $x\gets x_{new}$
   
 \quad \qquad $W\gets W_{new}$
   
  \quad\qquad $H\gets H_{new}$
   
  \qquad End if
  
 \quad End if
 
 $T\gets \lambda*T$
 
End while

Return $x$,$W$

The only nontrivial part is that the number and order of spins flipped are randomly generated, different from other reported schemes that flip only one spin at a time\cite{pi3,feng,ruan}. 
This is because flipping one spin at a time cannot efficiently ensure fast convergence for large-scale problems. Our scheme reduces the dependence of the search results on the initial values and improves the convergence speed noticeably. The effect of the number of flips per iteration on convergence rate was compared in Fig. \ref{fig:1}. Flipping one spin at a time evidently impedes the fast convergence. A fixed number of flips may also restrict the search space and consequently restrict to a local optimal solution. The figure clearly shows that the random flip is more conducive to achieving problem convergence and optimization.

\section{Experimental and simulated results}

\begin{table}[!t]
\footnotesize
\centering
\label{tab1}
\tabcolsep 11pt %space between two columns. ÓÃÓÚµ÷ÕûÁÐ¼ä¾à
\caption{\label{tab:table1}  Experimental, simulated, SG results for Max-Cut problems}
\begin{tabular}{c|cc|c|c|cc}
\hline
Density & \multicolumn{2}{c|}{Experiment} &Simulation &SG &\multicolumn{2}{c}{Increase rate (vs SG)} \\

 &max &ave & & &max &ave \\\hline

0.5 &153704302 &153039395 &116980988 &117136449	&31.22\% &30.65\% \\

0.6	&151121359 &150150495 &114841395 &114847656	&31.58\% &30.74\% \\

0.7	&157960371 &156723864 &119879172 &119899908 &31.74\% &30.71\% \\

0.8	&166267212 &164785788 &124132477 &126118964 &31.83\% &30.66\% \\

0.9 &181297365 &179571382 &134334306 &137372580 &31.97\% &30.72\% \\

1.0 &176701819 &175863542 &130222153 &117751430 &50.06\% &49.35\% \\

\hline
\end{tabular}
\end{table}

\begin{figure}[t]
\includegraphics[width=1\linewidth]{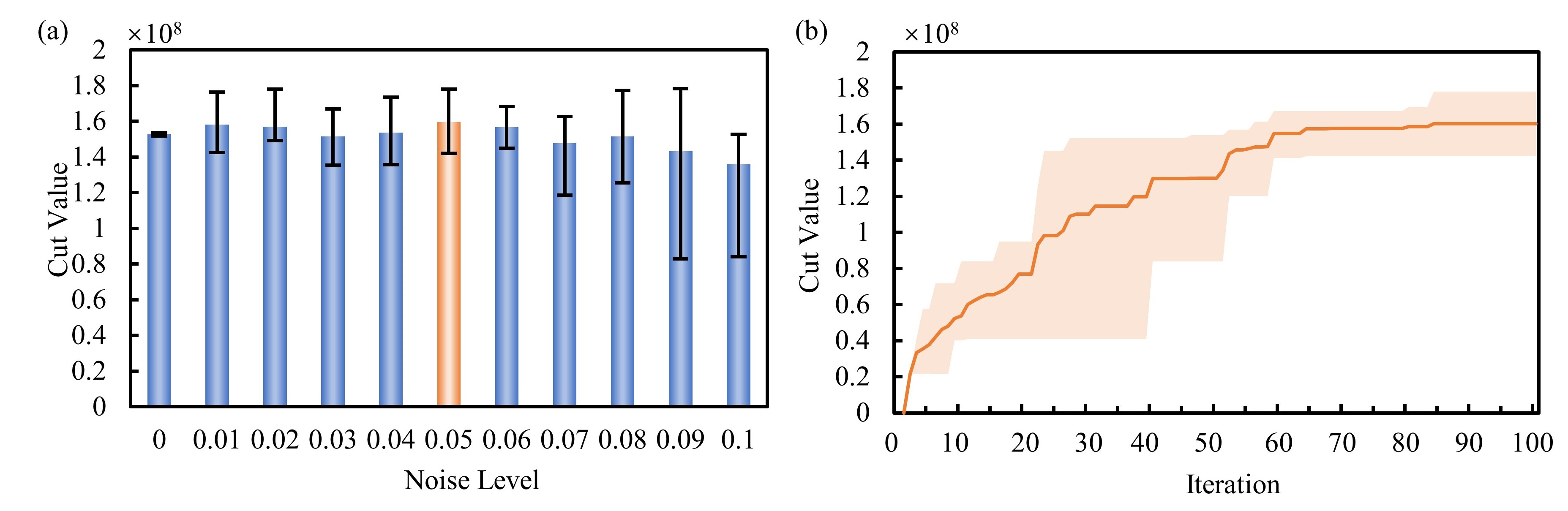}
\centering
\caption{
\textbf{The influence of injection noise on Q-SIM performance.} 
(a) Experimental results for 20736-node Max-cut problems (density=0.5) with digital noises in the level of 0-0.1. The result in the optimal noise level is marked as orange bar. 
(b) Experimental evolution of the cut value in the noise level of 0.05.The light orange region represents the interval of the result distribution from the ten experiments, with the mean indicated by the orange line.
}
\label{fig:noise}
\end{figure}

The instances of Max-Cut problem consist of graphs of 20736 nodes with random integer weights of floating-point numbers chosen from interval [-10,10] and the graph density $d\in\{0.5,1\}$, referring to Rudy-generated instances\cite{rudy}.  We applied the SG algorithm, numerical simulations and experiments to solve these problems, and the results are shown in Table \ref{tab:table1}.

The detection susceptible to noise may cause some uncertainty in experiments and we speculate that it is the discrepancy that makes it easier to jump out of the local optimum and fit better with the SA algorithm, resulting in a better solution. In fact, several related works have reported that noise-accelerated, noise-enhanced and noise-induced photonic Ising machines can be used to solve large-scale combinatorial optimization problems \cite{pi2,pris2,noisecim}. We add digital white noise to our experiments thereby accurately controlling the noise, the noise level being determined by the normalized noise variance. Taking the Max-Cut problem with a graph density of 0.5 as an example, Fig. \ref{fig:noise}(a) shows the results obtained by solving it using our Q-SIM under different noise levels. And the complete evolution of the cut value under optimal noise level in Fig. \ref{fig:noise}(a) is illustrated in Fig. \ref{fig:noise}(b).

\section{Discussion on the SIM performance of solving the Max-Cut problems }

\begin{table}[t!]
\footnotesize
\centering
\label{tab2}
\tabcolsep 11pt %space between two columns. ÓÃÓÚµ÷ÕûÁÐ¼ä¾à
\caption{\label{tab:table2}Performance comparison between our Q-SIM based phase-to-amplitude modulation based  and other approaches for solving Max-Cut problems.}
\begin{tabular}{ccccc}
\hline
Ising machine &Implementation &Problem Type &Problem Scale &Time to Resolution\\\hline

8-FPGA SB machine \cite{chip} &Easy &All-to-all,Weighted &16,384-node &1.2 ms \\

PRIS \cite{pris} &Easy &All-to-all,Unweighted &100-node &63 ns per-step\\

CIM with DOPO \cite{hon} &Very hard &All-to-all,Unweighted &100000-node &785 $\mu$s \\

CIM with OEPO \cite{li} &Hard &Sparse,Unweighted &56-node &4.5 $\mu$s \\

D-wave 2000Q \cite{dw,ptno}&Very hard &Sparse,Weighted &2500-node &$>10^{4}$ s (for 55 nodes)\\

Q-SIM (our method) &Easy &All-to-all,Weighted &20000-node &325 s \\
\hline
\end{tabular}
\end{table}
We compare the performance of the Q-SIM based phase-to-amplitude modulation in solving Max-Cut problems to other Ising machines in Table \ref{tab:table2}.
We evaluate relevant metrics, such as implementation, problem type and scale, time to resolution (or speed) and obtain the following conclusions:
\begin{enumerate}
\item Efficiently solving large-scale Max-Cut problems. Compared to most solutions \cite{chip,pris,li,dw}, we comfortably solve Max-Cut problems with size over 20,000, approaching the highest reported record so far \cite{hon}. In fact, we take the adjacent 10×10 pixels as an operation unit for the same encoding to ensure the consistency of the Ising system in our experiments, thus do not maximise the use of all pixel points. 
With further optimisation of the alignment and detection capabilities, it is feasible to scale up the problem hundredfold.

\item Flexible mapping of (non)fully connected Max-Cut problems with arbitrary amplitude. Considering experimental setups, many schemes prefer to demonstrate the process of solving the benchmark sparse unweighted Max-Cut problem \cite{pris,hon,li}. Obviously, being unweighted reduces the complexity, and fully connected problems are of more practical value and harder to implement than sparse ones \cite{chip}. As a result, many designs take great efforts to achieve fully connection. Quantum annealers sacrifice scale, and CIMs also address this deficiency by various schemes. And achieving weighted is even harder. However, it is where SIM excel and our design further magnifies this advantage by liberal switching between fully and non-fully connected, weighted and unweighted problems.

\item Simple and cost-effective experimental construction. Large power consumption and high costs are required by quantum annealers because of the cryogenic environment. Even CIMs impose rigorous experimental requirements.  Fiber oscillators of tens of kilometres are applied to keep optical loss and optical gain within thresholds and thus guarantee spin-to-spin coupling, bringing fairly large roundtrip loss. In contrast, our approach based on a simple SLM is superior in terms of experimental cost and manoeuvrability.
\end{enumerate}

Despite the fact that different Ising machines demonstrate their respective attractions in tackling Max-Cut problems, such as ultra-large scale \cite{hon}, ultra-high speed \cite{pris}, high stability \cite{li}, and arbitrary Max-Cut problem mapping \cite{chip,pris}, our design, by adopting a more economical experimental architecture, achieves the magnitude adjacent to the largest scale and free mapping of (non)fully connected, (un)weighted Max-Cut problems, which has greater practical implications for solving NP-hard problems. Although the design leaves much to be desired in terms of computational speed, which is constrained by the optoelectronic transmission of data and the refresh frequency of the SLM, it still exhibits speed advantages over electrical computation and even quantum annealing.

%%%%%%%%%%%%%%%%%%%%%%%%%%%%%%%%%%%%%%%%%%%%%%%%%%%%%%%
%%% Reference section. ²Î¿¼ÎÄÏ×
%%% citation in the content using "some words~\cite{1,2}".
%%% ~ is needed to make the reference number is on the same line with the word before it.
%%%%%%%%%%%%%%%%%%%%%%%%%%%%%%%%%%%%%%%%%%%%%%%%%%%%%%%

\end{appendix}
\end{document}